\documentclass[aps,prl,twocolumn,groupedaddress,showkeys]{revtex4-1}
\usepackage{graphicx}
\usepackage{dcolumn}
\usepackage{bm}
\usepackage{citesort}

\def \Re {{\rm Re}\, }

\def \be {\begin{equation}}
\def \ee {\end{equation}}
\def \ba {\begin{eqnarray}}
\def \ea {\end{eqnarray}}

\begin{document}

\title{Transformation optics with photonic band gap media}

\author{Yaroslav A. Urzhumov and David R. Smith}
\affiliation{Center for Metamaterials and Integrated Plasmonics, Pratt School of Engineering, Duke University, Durham, N. Carolina 27708, USA}
\email{yaroslav.urzhumov@duke.edu}

\date{\today}

\begin{abstract}
We introduce a class of optical media based on adiabatically modulated, dielectric-only, and potentially extremely low-loss, photonic crystals.
The media we describe represent a generalization of the eikonal limit of transformation optics (TO). The foundation of the concept is the possibility to fit frequency isosurfaces in the $k$-space of photonic crystals with elliptic surfaces, allowing them to mimic the dispersion relation of light in anisotropic effective media.
Photonic crystal cloaks and other TO devices operating at visible wavelengths can be constructed from optically transparent substances like glasses, whose attenuation coefficient can be as small as 10 dB/km, suggesting the TO design methodology can be applied to the development of optical devices not limited by the losses inherent to metal-based, passive metamaterials.

\end{abstract}

\keywords{Transformation Optics, Cloak, Eikonal Approximation, Geometrical Optics, Photonic Crystals, WKB Approximation.}

\maketitle


Transformation optics (TO) is a recently appreciated methodology for the design of novel electromagnetic devices~\cite{pendry_smith06}.
Devices created with the TO concept typically require exotic electromagnetic properties~\cite{pendry_ring99,smith_apl00},
which are unavailable in naturally occurring optical media such as solid-state crystals and amorphous substances (glasses).
In fact, the required properties are so difficult to achieve, that TO-based designs would have been dismissed prior to the development of mesoscopic ``metamaterial" fabrication methods~\cite{pendry_ring99,smith_apl00,schurig_smith06}. The desired material properties are usually available in metamaterials only at certain frequencies in the vicinity of a resonance band, which makes them inherently lossy.

The lossy nature of metamaterials has impeded progress in certain areas of metamaterial and TO research, including sub-diffraction limited imaging (superlensing)~\cite{pendrylens_prl00} and electromagnetic invisibility~\cite{pendry_smith06,schurig_smith06,hashemi_johnson10}. In the case of superlensing, which requires the exotic property of negative refractive index, an alternative, low-loss solution was proposed based on the existence of negative group velocity bands in photonic band gap media~\cite{luo_pendry_prb02,xiao_he04,schonbrun_summers06}. Photonic crystal-based imaging devices suffer from serious drawbacks, such as narrow operational bandwidth~\cite{luo_joannopoulos03} and inefficient coupling between free space and the crystal~\cite{xiao_he04,schonbrun_summers06,garciapomar_vesperinas09}. However, they offer the possibility of extremely low optical loss. So low the optical attenuation in a photonic band gap medium may be that it could span millions of wavelengths and yet be virtually transparent; the glass used in optical fibers~\cite{smith72} is a famous example of such a low-loss dielectric component.

Transformation optics designs result in inherently complex media, typically requiring the exotic property of superluminal phase velocity (i.e., refractive index $n<1$) combined with anisotropy (such as in beam shifters~\cite{rahm_smith08,garciapomar_vesperinas09}), spatial gradients in refractive index (as in the flattened Luneburg lens~\cite{kundtz_smith10}), or both anisotropy and spatial gradients (as in the cloak of invisibility~\cite{pendry_smith06,schurig_smith06}). The ``invisibility cloak" in particular has stimulated considerable interest over the past several years, and serves as a natural test bed for various TO-inspired concepts and sophisticated metamaterial designs. In this article, we similarly design a cloak to demonstrate the general ability to manipulate light propagation using photonic crystals, mimicking TO media.

In its full generality, TO is able to map essentially any volume with an arbitrary bounding surface onto another arbitrarily shaped volume. For an invisibility cloak, one can restrict these transformations to radially symmetric maps.
In two dimensions, a ``perfect" cloak can be obtained from a coordinate transformation $r'=q(r)$
that maps an annulus $a<r<b$ onto the disk $0<r'<b$~\cite{pendry_smith06,schurig_smith06}.
To obtain a two-dimensional cloak, the constitutive parameters in the physical space (with radial coordinate $r$) 
should be chosen as follows~\cite{schurig_smith06,cai_milton07}:
\begin{eqnarray}
\epsilon_r = \mu_r = (q/r)/q',  \epsilon_\theta=\mu_\theta = 1/\epsilon_r, \epsilon_z=\mu_z = q'(q/r),
\label{eq:perfect_cloak_2d}
\end{eqnarray}
where $q'\equiv dq(r)/dr$.
The refractive indices corresponding to radial ($n_r$) and azimuthal ($n_\theta$) propagation in the perfect cloak~(\ref{eq:perfect_cloak_2d}) are
$n_r = q'$ and $n_\theta = q/r$, respectively, for both TE~\cite{schurig_smith06} and TM~\cite{cai_shalaev07,smolyaninov_shalaev09} polarizations.

\begin{figure}
\centering
\includegraphics[width=0.5\columnwidth]{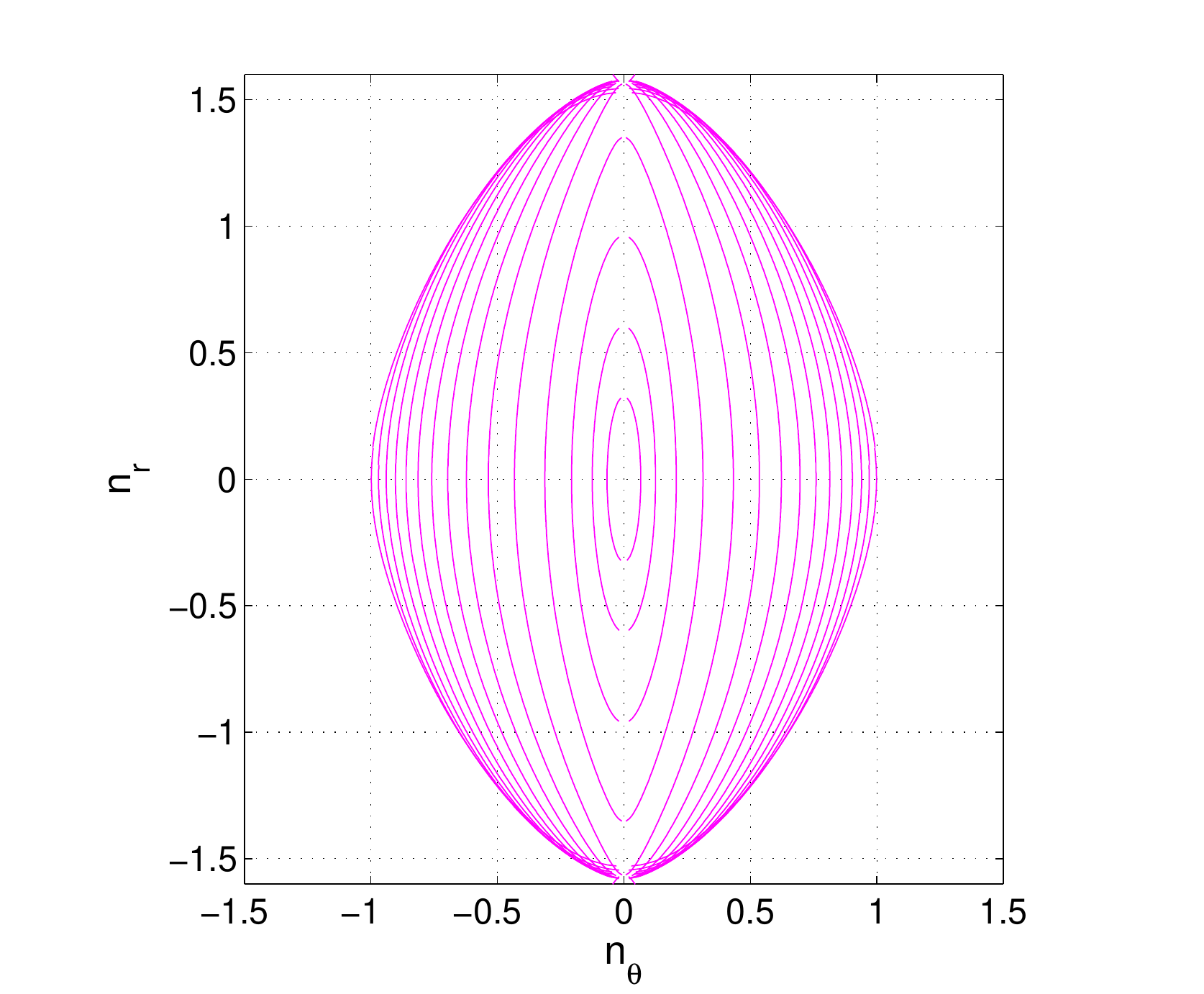}
\caption{
(Color online)
Dispersion relation for TM waves in a one-dimensional bi-layer photonic crystal
with dielectric permittivities $\epsilon_1=12$, $\epsilon_2=1$ and filling fraction $t_1/t=0.9$.
Horizontal axis is labeled by refractive index $n_\theta=k_\theta/k_0$ corresponding to propagation in the periodic direction of the crystal, while the vertical axis is refractive index in the other direction ($n_r=k_r/k_0$).
}
\label{fig:elliptic_isocontours}
\end{figure}

In the geometrical optics limit one may abandon the notion of $\epsilon$ and $\mu$ in favor of the refractive index, since only the optical path length has relevance.
The refractive index is defined as the wavenumber $k$ corresponding to wave propagation in a certain direction, normalized to the vacuum wavenumber $k_0=\omega/c$, i.e. $n(\vec k)={|\vec k|}/{k_0}$.
Knowing the function $n(\vec k)$, or $\omega(\vec k)$--the dispersion relation--is sufficient to describe the dynamics of light in transparent media~\cite{russell_birks99}.
Utilizing the dispersion relation to describe wave propagation is more general than effective medium descriptions based on the tensors $\epsilon,\mu$, since the properties of band gap optical media where propagating Bloch-Floquet modes exist, such as gratings and photonic crystals, can readily be analyzed~\cite{joannopoulos_winn95,notomi00}.
If the function $n(\vec k, \vec r)$ changes slowly (adiabatically) with position $\vec r$,
i.e. $\left|\nabla_{\vec r} n(\vec k, \vec r)\right|\ll \frac{1}{\lambda}$,
light propagation can be described in the WKB, or semiclassical, approximation~\cite{landau_lifshitz_qm76,russell_birks99}.

The adiabaticity of $n(\vec k, \vec r)$ implies that it is continuous everywhere in space.
Where a discontinuity exists, however, the function $n(\vec k, \vec r)$ is not sufficient: it cannot describe transmission and reflection at the discontinuity, a limitation that applies to the sharp transitions between a photonic crystal and free space.
Whereas an effective medium description would enable the reflectance and transmittance to be calculated from an effective wave impedance $Z=\sqrt{\mu\epsilon^{-1}}$,
the wave impedance is generally an ill-defined concept for photonic crystals. The lack of a well-defined wave impedance in photonic crystals leads to the generation of multiple refracted and reflected beams (Bragg diffraction), and a dependence of reflectivity upon the microscopic details of the surface~\cite{xiao_he04,frei_johnson05}. 
Recent studies of negatively refracting photonic crystals with regards to their imaging properties show that improving the coupling between free-space and crystal modes is possible~\cite{xiao_he04,frei_johnson05,schonbrun_summers06}, although very difficult due to lack of a general theoretical model for the coupling. We note that the mode coupling problem is essentially unrelated with the dispersion engineering problem that we address here.

Once the effective medium description is replaced by the $\omega(\vec k, \vec r)$ description,
two features potentially useful for optical TO media can be noted.
{\it Firstly}, materials with permittivity or permeability less than one, which are necessarily dispersive and {\it lossy}, are no longer needed.
The dispersion relation of a photonic crystal near the $\Gamma$-point ($\vec k=0$) can be written as a Taylor expansion with even powers of $k$:
\be \omega^2(\vec k)=\omega_\Gamma^2 + v_r^2{k_r^2} + v_\theta^2{k_\theta^2} + O(k^4),
\label{eq:quadratic-dispersion}
\ee
or, equivalently,
$ \omega^2(\vec k)/c_0^2=\frac{k_r^2}{n_r^2} + \frac{k_\theta^2}{n_\theta^2} + O(k^4)$,
where $c_0$ is the speed of light in vacuum and
$n_{r,\theta}^2 = \frac{\omega^2-\omega_\Gamma^2}{\omega^2}\frac{c_0^2}{v_{r,\theta}^2}$.
In the above we assumed that the propagation band has a positive group velocity, and thus, $\omega>\omega_\Gamma$;
similar formulas for a negative-refraction band can be obtained by changing the sign of $v_{r,\theta}^2$ in the expressions above. The semi-axes of the elliptical contours of equal frequency in the $\vec k$-space are given by
$k_r^{\max}=n_r \omega/c_0$ and $k_\theta^{\max}=n_\theta \omega/c_0$.
In the $\Gamma$-point ($\omega=\omega_\Gamma$), $n_\theta=n_r=0$. Note that the effective indexes $n_{\theta,r}$ are no longer related to any effective medium theory and, consequently, not restricted by Voigt-Reiss or Hashin-Shtrikman bounds arising from the classical homogenization theories~\cite{zhikov94}. {\it Secondly}, anisotropy of the refractive index is possible in dielectric-only crystals, and not only for TM waves,
but also for TE waves. This contrasts with the behavior of TE waves in dielectric effective media, whose refractive index $n=\sqrt{\epsilon_z}$ is always isotropic.



While the approximately elliptic equal frequency contours (\ref{eq:quadratic-dispersion}) are very common in the band structure of photonic crystals in the vicinity of $\Gamma$-points ($k=0$)~\cite{joannopoulos_winn95,notomi00,luo_pendry_prb02,luo_joannopoulos03},
some bands preserve their ellipticity throughout a relatively wide band.
The existence of such an elliptic band in a one-dimensional photonic crystal formed from two alternating dielectric layers is illustrated by Fig.~\ref{fig:elliptic_isocontours}.
The dispersion relation of the bi-layer crystal is known in analytical form~\cite{kronig_penney31}.
Similar elliptic bands can be found in higher-dimensional photonic crystals, whose dispersion relation can be computed numerically.



Theoretically, transformation optics is a powerful framework that enables unprecedented control over the propagation of electromagnetic waves.
Anisotropy of the material properties implementing non-trivial coordinate transformations
is a typical, and almost inevitable, property of the TO-based designs.
A relatively simple TO-based device that uses anisotropic media is a beam shifter~\cite{rahm_smith08}.
Its simplicity is due to the fact that a transverse shift of the beam can be obtained with a piecewise linear transformation,
which results in piecewise constant material properties.
In other words, the linear beam shifter can be implemented as a flat slab of a medium with uniform, although anisotropic, effective index.
Previously, reflectionless beam shifters based on embedded coordinate transformations have been demonstrated theoretically in the effective medium regime~\cite{rahm_smith08}.

A linear beam shifter with the shift parameter $a=\tan(\phi)$ deflects the beam at an angle $\phi$, resulting in a transverse shift $\Delta y=ad$, where $d$ is the thickness of the shifter~\cite{rahm_smith08}. Below, we design and simulate a beam shifter operating in a higher Bloch-Floquet band of a photonic crystal. A bi-layer crystal illustrated and described in Fig.~\ref{fig:elliptic_isocontours} is selected for this demonstration. The operational frequency is chosen such that the two refractive indexes of the beam shifter medium, $n_{1,2}=\left(1+a^2/2(1\mp\sqrt(1+4/a^2))\right)^{1/2}$, match the two principal indexes of the photonic crystal; this leads to the choice $k_0t=2.449$, where $t$ is the crystal periodicity. The resulting beam shifter has a shift parameter of $a=0.6187$ and shift angle $\phi=31.7^\circ$. The principal values of refractive index are $n_1=0.7374$ and $n_2=1.3561$. The principal direction corresponding to smaller index $n_1$ makes an angle $\theta=36^\circ$ with the horizontal axis; the layers themselves make the same angle $\theta$ with the vertical axis.

\begin{figure}
\centering
\begin{tabular}{cc}
\multicolumn{2}{c}{\includegraphics[width=0.3\columnwidth]{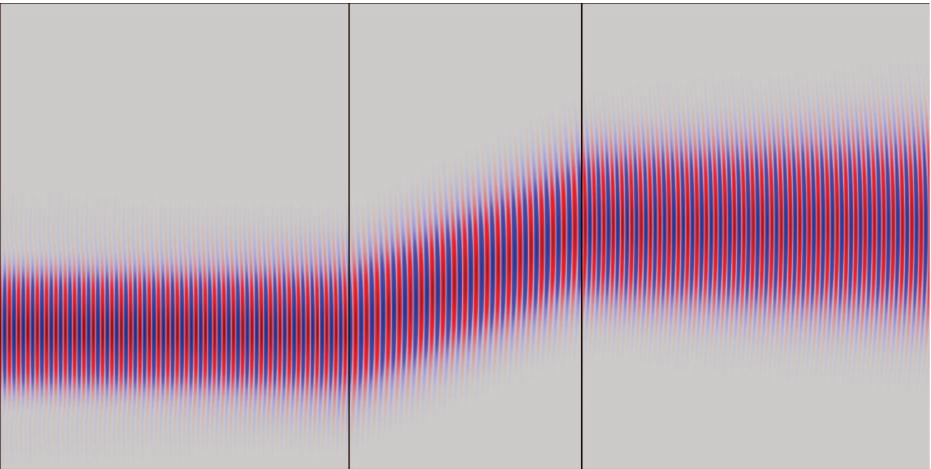}} \\
\multicolumn{2}{c}{(a)}\\
\includegraphics[width=0.25\columnwidth]{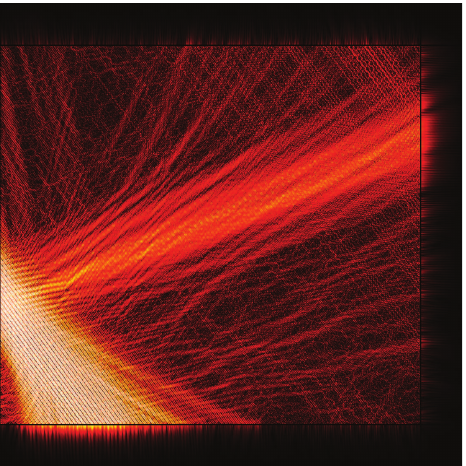}
&\includegraphics[width=0.25\columnwidth]{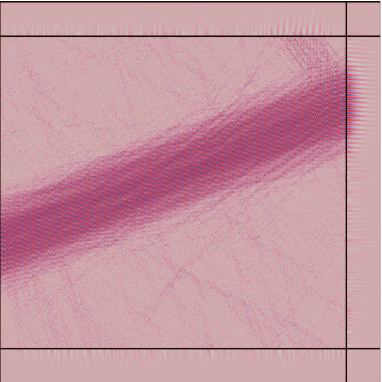}\\
(b)&(c)\\
\end{tabular}
\caption{
(Color online) Parallel-slab beam shifter based on embedded coordinate transformations.
(a) Gaussian beam incident from vacuum on a slab of anisotropic effective medium. Color shows out-of-plane magnetic field $\Re(H_z)$.
(b) Gaussian beam incident from vacuum on a one-dimensional photonic crystal.
Color shows $|H_z|$.
(c) Same as (b), except the Gaussian excitation is projected onto a single Bloch-Floquet mode of the photonic crystal. Color shows $\Re(H_z)$.
}
\label{fig:beam_shifter}
\end{figure}

The propagation of a Gaussian beam incident on a slab of homogeneous anisotropic medium with the refractive indexes described above is shown in Fig.~\ref{fig:beam_shifter}(a). The beam traveling horizontally in free space propagates at an angle $\phi=31.7^\circ$ inside the shifter.
Note that the phase fronts remain vertical inside the shifter, indicating that the directions of the phase and group velocities differ. Fig.~\ref{fig:beam_shifter}(b) shows the propagation of a beam with the same parameters incident on a photonic crystal slab with the same effective indexes.
The crystal slab is terminated with perfectly matched layers to reduce internal reflections.
We observe that a free-space Gaussian mode excites two beams inside the crystal, indicating that its frequency lies within a higher propagation band.
The upward-refracted beam follows the same path as the beam in Fig.~\ref{fig:beam_shifter}(a). The downward-refracted beam is caused by Bragg diffraction; multiple solutions for the refracted wavenumbers can be understood as multiple intersections of a straight line $k_y=0$ representing conservation of tangential wavenumber $k_y$ across the free space-crystal interface, with a periodically repeating pattern of isofrequency ellipses in the $k_x-k_y$ space.
To provide a clearer demonstration of the concept, we excite the crystal slab with a single mode of the crystal, modulated with the same Gaussian envelope, as shown in Fig.~\ref{fig:beam_shifter}(c). The mode has no phase shift in the vertical direction, which corresponds to a horizontally incident beam. As expected, the second (diffracted) beam is eliminated while the first (refracted) beam is intact. Notably, the phase fronts of the Bloch modes in Fig.~\ref{fig:beam_shifter}(c) remain vertical, in perfect agreement with effective medium simulation in Fig.~\ref{fig:beam_shifter}(a).

\begin{figure}
\centering
\begin{tabular}{cc}
\includegraphics[width=0.45\columnwidth]{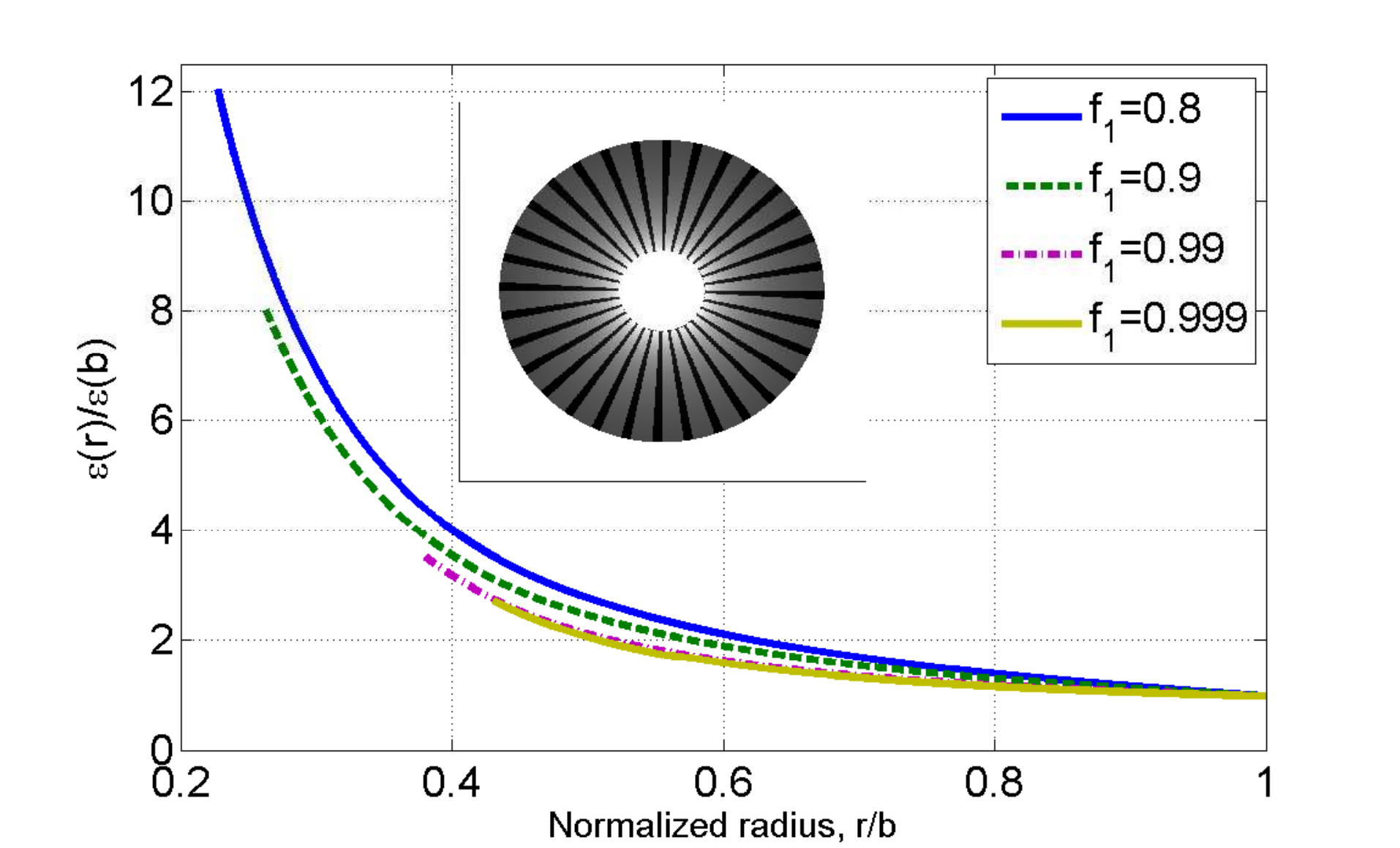}
&\includegraphics[width=0.45\columnwidth]{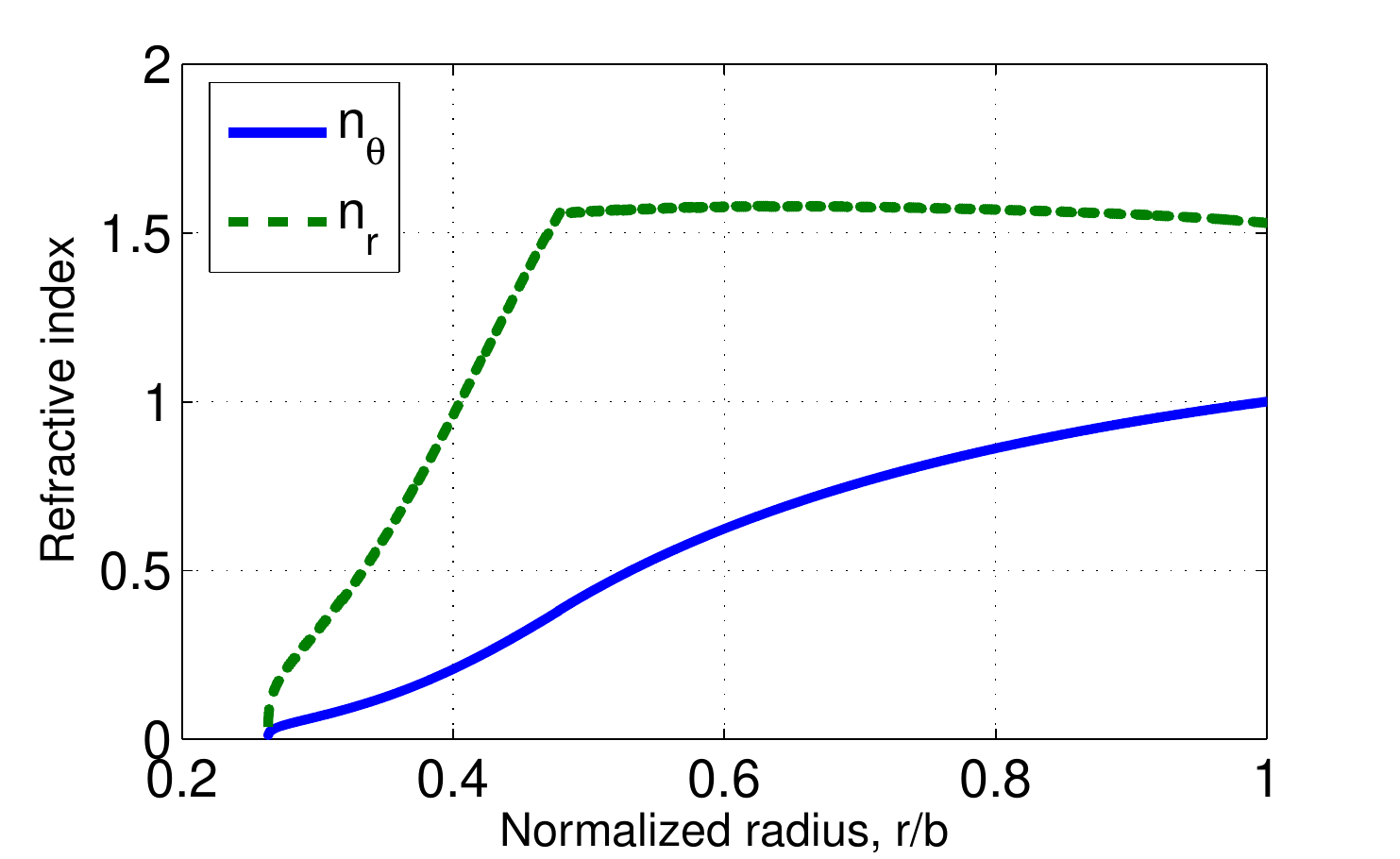}\\
(a)&(b)\\
\end{tabular}
\caption{(Color online)
(a)  Scaling of the dielectric constants with radius,
for the photonic crystal cloaks with $\epsilon_1/\epsilon_2=12$ (TM polarization).
The inset shows the geometry of the cloak.
(b) Refractive index distribution in the photonic crystal cloak with dielectric contrast $\epsilon_1/\epsilon_2=12$ and filling fraction $t_1/t=0.9$.
}
\label{fig:cloak_indexes}
\end{figure}

To achieve a more complex TO design, such as a cloak, we can add an adiabatic spatial modulation to the one-dimensional crystal.
The cylindrical cloak of invisibility derived from equations (\ref{eq:perfect_cloak_2d}) requires $n_\theta<n_r$.
We have found numerically that in all elliptic bands of the bi-layer photonic crystal (except the lowest, acoustic branch), the refractive index for the propagation along the layers is always greater than in the other direction: $n_\perp<n_{||}$; this was found for both TM and TE polarizations. Note that this property of higher-order bands is opposite to the behavior of the acoustic TM band,
which has $\epsilon_\perp = \langle \epsilon^{-1} \rangle^{-1} \le \epsilon_{||}=\langle \epsilon \rangle$ and therefore
$n_\perp=\sqrt{\epsilon_{||}} \ge n_{||}=\sqrt{\epsilon_\perp}$,
as dictated by the Voigt-Reiss inequalities of homogenization theory~\cite{zhikov94}.
This finding mandates that the crystal layers must be oriented radially to achieve the required index distribution of the cloak, as shown in the inset to Fig.~\ref{fig:cloak_indexes}(a).


The procedure for generating a radial transformation for eikonal cloaks has been introduced recently in Ref.~\cite{urzhumov_smith_njp10}.
In this case, it must be amended to account for the additional geometric constraint.
In contrast with effective medium-based cloaks, the unit cells in photonic crystal cloaks must change continuously between layers.
In a rotationally-invariant design, to preserve the number of unit cells per circumference,
unit cell size in the azimuthal direction, $t$, must change linearly with radius:
$ t(r)=t_b \frac{r}{b}$.
To design a cloaking transformation, we assume that, besides $t(r)$, there is one more parameter, $p(r)$, that affects the band structure.
Potential choices of $p(r)$ in a bi-layer photonic crystal include (A) the dielectric contrast between the two components, $\epsilon_1/\epsilon_2$,
(B) the filling fraction, $t_1/t$, or (C) the scaling factor for both permittivities, $\alpha$, such that
$\epsilon_{1,2}(r)=\epsilon_{1,2}(b) \alpha(r)$, where $\alpha(b)=1$.
To find the function $p(r)$ that gives an eikonal cloak, one may start from the differential equation of Ref.~\cite{urzhumov_smith_njp10},
$ \frac{dr}{r}=\frac{dn_\theta}{n_r(p,t)-n_\theta(p,t)}$,
where $n_{r,\theta}$ are known functions of $p$ and $t$ calculated from the dispersion relation. 
Assuming a linear function for $t(r)$, the unknown function $p(r)$ is found by integrating a first-order ODE
$
\frac{dp}{dr}=
\left(
\frac{n_r(p,t(r))-n_\theta(p,t(r))}{r} - \frac{\partial n_\theta(p,t(r))}{\partial t} \frac{dt}{dr}
\right)\left(\frac{\partial n_\theta(p,t(r))}{\partial p}\right)^{-1}\equiv F(p,r)$.
The choice (C) allows an analytical simplification of the dispersion engineering problem.
When both the dielectric contrast and the filling ratio are constant, the band structure is a function of a single parameter, $\beta(r)=t(r)\alpha^{1/2}(r)$, and the transformation can be found as a simple quadrature: $\int\frac{dr}{r}=\int\frac{d\beta}{n_r(\beta)-n_\theta(\beta)}\frac{d n_\theta}{d\beta}$.
The resulting cloaking transformation and permittivity distribution for photonic crystals with radius-independent dielectric contrast $\epsilon_1/\epsilon_2=12$ and several choices of the filling ratio are plotted in Fig.~\ref{fig:cloak_indexes}(a). In the remainder of this article, we will use the filling fraction $t_1/t=0.9$, which leads to an eikonal cloak with aspect ratio $a/b=0.2631$. Fig.~\ref{fig:cloak_indexes}(b) shows the distribution of refractive indexes in such a cloak. 


\begin{figure}
\centering
\begin{tabular}{cc}
\includegraphics[width=0.35\columnwidth]{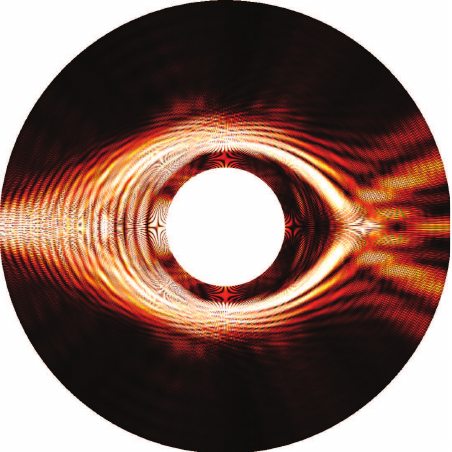}
&\includegraphics[width=0.35\columnwidth]{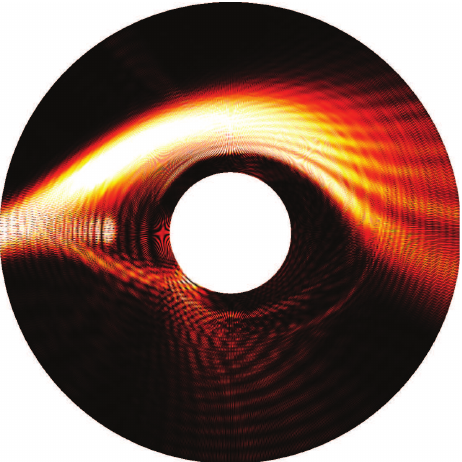}\\
(a)&(b)\\
\end{tabular}
\caption{(Color online)
(a) Cloaking of a normally incident Gaussian beam modulated with the desired mode profile of the photonic crystal.
(b) Same as (a), but with the angle of incidence $\theta_{inc}=15^\circ$.
}
\label{fig:cloak_gaussian}
\end{figure}

The operation of the photonic crystal cloak is demonstrated using electromagnetic simulations
performed with COMSOL, a Finite Element Method solver.
The geometry of the cloak consists of two concentric circles of radii $b=0.4$ (m) and $a=0.10524$ (m).
The inner disk represents the cloaked object, on which perfect electric conductor boundary conditions are imposed.
The space between the circles is filled with a quasi-periodic photonic crystal with $N=512$ radially converging layers.
In order to suppress additional beams, a single mode of the crystal is injected in the simulation in the same fashion as was done in the beam shifter model in Fig.~\ref{fig:beam_shifter}(c). As seen in Fig.~\ref{fig:cloak_gaussian}(a), a Gaussian-like beam is created in the crystal. The beam begins to diverge as it travels towards the center; then it bends around the cloaked object and re-forms as another beam on the opposite side of the cloaking device. Excitation of the same mode with a non-zero phase shift per unit cell results in a Gaussian beam traveling at an angle, as shown in Fig.~\ref{fig:cloak_gaussian}(b).

In conclusion, we have demonstrated theoretically and in high-resolution electromagnetic simulations that the concepts of transformation optics can be implemented with Bloch-Floquet waves in adiabatically modulated photonic crystals. This includes a demonstration of a dielectric-only optical cloak, whose practical implementation can be composed of materials that are macroscopically transparent in the entire visible spectrum. This paradigm creates hope for a new class of optical devices capable of being scaled to arbitrarily large dimensions.


This work was partially supported through a Multiple University Research Initiative,
sponsored by the U.S. Army Research Office (Contract No. W911NF-09-1-0539).

\bibliographystyle{unsrt}
\bibliography{cloaking}

\end{document}